\documentclass{ws-mpla}
\usepackage[super]{cite}
\usepackage{graphicx}
\usepackage{amsmath}
\usepackage{bm}

\begin{document}


\title{Angular decorrelations in
 $\gamma + 2 jets$ events at high
energies in the Parton Reggeization Approach}

\author{Anton Karpishkov \footnote{karpishkoff@gmail.com}, Vladimir Saleev \footnote{saleev@samsu.ru} }

\address{Department of Physics, Samara National Research University, Moskovskoe Shosse,
34, Samara, 443086, Russia}

\author{Alexandra Shipilova \footnote{alexshipilova@samsu.ru}}
\address{Department of Physics, Samara National Research University, Moskovskoe Shosse,
34, Samara, 443086, Russia\\ and \\Joint Institute for Nuclear
Research, Dubna 141980, Russia}

\maketitle

\begin{history}
\received{Day Month Year} \revised{Day Month Year}
\end{history}

\begin{abstract}
  We study associated production of prompt photon and
two jets at  high energies in the framework of the parton
Reggeization approach, which is based on multi-Regge factorization
of hard processes and Lipatov's effective theory of Reggeized gluons
and quarks. In this approach, initial-state off-shell effects and
transverse momenta of initial partons  are included in a
gauge-invariant way. We compute azimuthal angle difference spectra
in $\gamma + 2 jets$ events and compare results with data from D0
Collaboration at the Tevatron. It was found that agreement between
predictions and data can be achieved only under assumption on strong
violation in transverse momentum ordering during initial-state QCD
evolution which corresponds multi-Regge kinematical regime instead
of DGLAP picture.
\end{abstract}


\section*{Introduction}

 Theoretical and experimental study of associated production of prompt photon
 and jets with large transverse momenta in high-energy hadronic collisions is a very important task for various reasons.
First, this is a challenging test of our understanding of
higher-order corrections in quantum chromodynamics (QCD). In
general, it is a nontrivial task to provide reliable predictions for
multiscale and correlational observables, based on the conventional
Collinear Parton Model (CPM) of QCD. Second, correlational
observables in photon plus jets production are primary tools in
experimental searches of Double Parton Scattering(DPS) mechanism
manifestations  in hadron-hadron collisions at high energies
\cite{DPS}.

In the present paper we discuss associated production of prompt
photon and two jets in $p\bar p$ collisions at the $\sqrt{S} = 1.96$
TeV which has been observed by the D0 Collaboration~\cite{D0gamma2}.
A comparison of theoretical predictions obtained in the
next-to-leading order (NLO) approximation of CPM with the measured
cross sections  and the azimuthal angle difference distributions
pointed towards DPS as the reliable source of events in the region
of small $\Delta\phi$ \cite{D0gamma2}.

However, there are many evidences that $k_T-$factorization approach
of high-energy QCD  is more adequate for description of multi-jet
production processes, specially for decorrelation observables such
as azimuthal angle difference spectra or rapidity difference spectra
~\cite{NSS2013,gammaHERA}. In the $k_T-$factorization approach the
main part of high-order corrections from hard real partons emission
is included  using the unintegrated Parton Distribution Functions
(unPDFs). The presence of non-vanishing transverse momenta of
initial-state partons before hard interaction provides a flat
behavior of photon-jets azimuthal angle difference spectra in the
region far from back-to-back configuration, together with DPS
contribution.

In Ref.~\cite{D0gamma2}, the study of azimuthal correlations in
$\gamma+2 jets$ events is introduced to be an extremely sensitive
tool for DPS signals search. They measured azimuthal angle
difference $\Delta \phi_A$ between sum of photon and leading jet
transverse momenta ${\bf p}_{T}^A={\bf p}_T^{\gamma}+{\bf
p}_{T}^{jet1}$ and transverse momentum of second (sub-leading) jet
${\bf p}_{T}^{jet2}$. Such a way, in the LO CPM we should expect a
strong peak near $\Delta \phi=\pi$, and broadening of this peak
should be explained by emission of additional hard jet in the NLO
approximation of the CPM. Such calculations, performed using
SHERPA~\cite{SHERPA} program (see Figs. 9-11, in
Ref.~\cite{D0gamma2}), fail to describe data from D0 Collaboration
at $\Delta\phi<\pi/4$ without inclusion of sufficient contribution
from DPS mechanism.

Here, we examine the Single Parton Scattering (SPS) mechanism to
describe data on $\gamma +2 jets$ associated production, when
higher-order QCD corrections are partially taken into account
using the parton Reggeization approach (PRA)~\cite{NSS2013,NKS2017}.

\section{Parton Reggeization Approach}
\label{PRA}
 The master formulas of Leading-Order (LO) approximation of PRA are presented below.
The more detailed description can be found in \cite{NKS2017}, while
the development of PRA in the NLO approximation is further discussed
in \cite{PRANLO1}. The main ingredients of PRA are the factorization
formula for hard processes in the Multi-Regge Kinematics (MRK), the
Kimber-Martin-Ryskin (KMR) unPDFs~\cite{KMR} and the
gauge-invariant amplitudes of hard processes with off-shell
initial-state partons, derived using the Lipatov's Effective Field
Theory (EFT) of Reggeized gluons~\cite{Lipatov95} and Reggeized
quarks~\cite{LipatovVyazovsky}.

The factorization formula of PRA in LO approximation can be obtained
from the relevant factorization formula of the CPM~\cite{NKS2017}
for the auxiliary hard subprocess with two additional final-state
partons using the modified Multi-Regge Kinematics (mMRK)
approximation. Last one correctly reproduces the multi-Regge and
collinear limits of corresponding QCD amplitude. In this
mMRK-approximation one has:
  \begin{equation}
  d\sigma = \int \frac{dx_1}{x_1} \int \frac{d^2{\bf q}_{T1}}{\pi} {\Phi}_1(x_1,t_1,\mu^2)
\int \frac{dx_2}{x_2} \int \frac{d^2{\bf q}_{T2}}{\pi}
{\Phi}_2(x_2,t_2,\mu^2)\cdot d\hat{\sigma}^{\rm PRA},
\label{eqI:kT_fact}
  \end{equation}
where $t_{1,2}={\bf q}_{T1,2}^2$, the partonic cross-section
$\hat\sigma^{\rm PRA}$ in PRA is determined by squared PRA
amplitude, $\overline{|{\cal A}_{PRA}|^2}$. Despite the fact that
four-momenta ($q_{1,2}$) of partons in the initial state of ${\cal
A}_{PRA}$ are off-shell ($q_{1,2}^2=-t_{1,2}<0$), the PRA
hard-scattering amplitude is gauge-invariant because the
initial-state off-shell partons are treated as Reggeized gluons
($R$) or Reggeized quarks ($Q$) in a sense of gauge-invariant EFT
for QCD processes in MRK, introduced by L.N. Lipatov
in~\cite{Lipatov95}. The Feynman rules of this EFT are written down
in Ref.~\cite{AntonovLipatov,LipatovVyazovsky}.

The unPDF in PRA formally coincides with Kimber-Martin-Ryskin (KMR)
unPDF~\cite{KMR} and can be presented as follows:
  \begin{equation}
  \Phi_i(x,t,\mu^2) = \frac{T_i(t,\mu^2)}{t} \frac{\alpha_s(t)}{2\pi} \sum_{j=q,\bar{q},g}
   \int\limits_x^{1-\Delta_{KMR}} dz\ P_{ij}(z)\cdot \frac{x}{z}f_{j}\left(\frac{x}{z},t \right)
   , \label{eqI:KMR}
  \end{equation}
where $\Delta_{KMR}(t,\mu^2)=\sqrt{t}/(\sqrt{\mu^2}+\sqrt{t})$ is
the KMR cutoff function~\cite{KMR}, which regularizes infrared (IR)
divergence at $z_{1,2}\to 1$ and introduces rapidity-ordering
between initial-state and final-state partons. The collinear
singularity in KMR unPDF is regularized by the Sudakov formfactor:
  \begin{equation}
  T_i(t,\mu^2)=\exp\left[ - \int\limits_t^{\mu^2} \frac{dt'}{t'} \frac{\alpha_s(t')}{2\pi}
\sum\limits_{j=q,\bar{q},g} \int\limits_0^{1-\Delta_{KMR}} dz\
z\cdot P_{ji}(z) \right], \label{eq:Sudakov}
  \end{equation}
which resums doubly-logarithmic corrections $\sim\log^2 (t/\mu^2)$
in the leading-logarithmic approximation.

In contrast to most of studies in the $k_T$-factorization, the
gauge-invariant matrix elements with off-shell initial-state partons
(Reggeized quarks and Reggeized gluons) of Lipatov's
EFT~\cite{Lipatov95, LipatovVyazovsky} allow one to study arbitrary
processes involving non-Abelian structure of QCD without violation
of Slavnov-Taylor identities due to the nonzero virtuality of
initial-state partons. This approach, together with KMR unPDFs gives
stable and consistent results in a wide range of phenomenological
applications, which include the description of the different spectra
of single jet and prompt-photon inclusive production
\cite{gamma,gamma-jet}, two jets~\cite{NSS2013} or two
photons~\cite{Diphotons} in $pp$ and $p\bar p$ collisions, and
photon plus jet in $\gamma p$ collisions at HERA
Collider~\cite{gammaHERA}.

\section{Associated production of photon and two jets in PRA}
\label{spectra}

To describe experimental data of prompt photon spectra we should
take into account two production mechanisms. The first one is a
direct production, when photons are produced in hard quark-gluon
collisions. The second one is a fragmentation production, when
quarks or gluons produced in hard collisions  emit collinear
photons. Fragmentation production of prompt photons can be strongly
suppressed by experimental cuts, producing only a few percents from
total number of events at small photon transverse momentum. This
fact allows us to neglect here the small fragmentation contribution.

\begin{figure}[h]
\centering
\includegraphics[width=0.8\textwidth, angle=0]{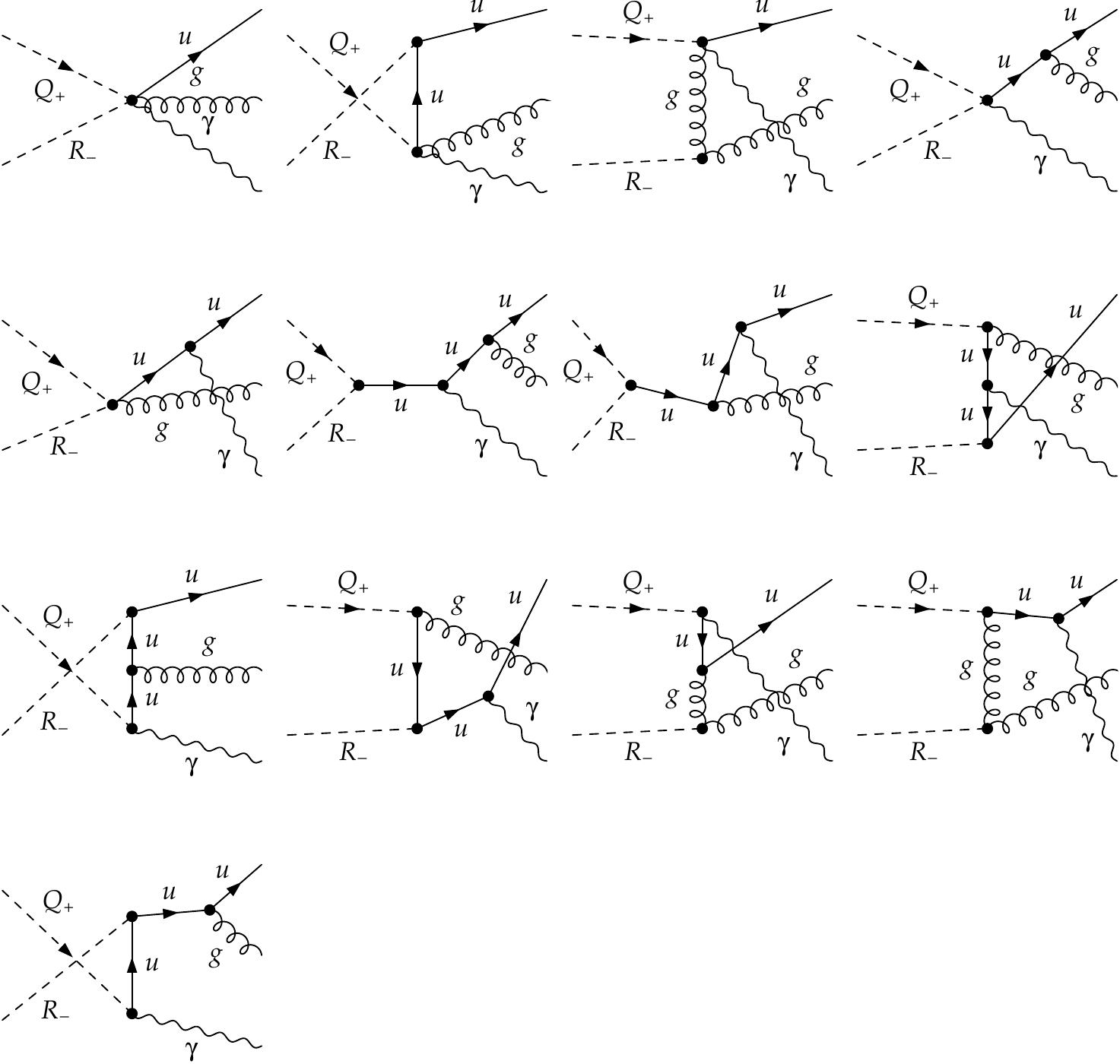}
\caption{Feynman diagrams for process $RQ\to g q \gamma$ in
Lipatov's EFT.}
\label{fig-0}       
\end{figure}

The LO PRA processes which contribute in direct $\gamma+2 jets$
events are the following:
\begin{eqnarray}
Q+R &\to& q + g +\gamma, \label{QR}\\
R+R &\to& q + \bar q+ \gamma, \label{RR}\\
Q + \bar Q &\to& q(q') + \bar q (\bar q')+ \gamma,\\
Q + \bar Q &\to&
g + g + \gamma,\\
Q + Q&\to& q + q +\gamma, \label{QQ}\\
Q + Q'&\to& q + q' +\gamma \label{QQ1} \label{QaQ}.
\end{eqnarray}
We omit  the processes, which contributions are smaller than 1 \% in
total cross section. The main contribution comes from the process
(\ref{QR}), which is described by the set of Feynman diagrams of
Lipatov's EFT presented in the Fig.~\ref{fig-0}.

The amplitudes of all {above-mentioned} LO PRA processes can be
obtained in analytical form using model-file \textbf{ReggeQCD}
\cite{ReggeQCD}, which implements the Feynman rules of Lipatov's EFT
in \textbf{FeynArts}~\cite{FeynArts} at tree level. To generate the
gluon, $\Phi_g(x,t,\mu^2)$, and quark, $\Phi_q(x,t,\mu^2)$, unPDFs,
according to the Eq.~(\ref{eqI:KMR}) we use the LO PDFs from the
Martin-Roberts-Stirling-Thorne (MRST) set~\cite{MRST}.

We set the renormalization and factorization scales equal to the
transverse momentum of leading jet, $p_{1T}$: $\mu_R=\mu_F=\xi
p_{1T}$, where $\xi=1$ for the central lines of our predictions, and
we vary $1/2<\xi<2$ to estimate the scale uncertainty  of our
prediction, which is shown in the figures by the gray band.

After our numerical calculations, based on analytical amplitudes
obtained from the Lipatov's EFT, have been completed, we have got an
opportunity to perform a cross-check of our results with MC
generator {\it Katie} ~\cite{Katie}.

The last one uses gauge-invariant scattering
amplitudes with off-shell initial-state partons, obtained using the
spinor-helicity techniques and BCFW-like recursion relations for
such amplitudes~\cite{Hameren1,Hameren2}. This formalism for
numerical generation of off-shell amplitudes is equivalent to the
Lipatov's EFT at the tree level.

\section{DGLAP and MRK regimes}
Now, we come back to the factorization formula (\ref{eqI:kT_fact})
where upper limits of integrals over $t_{1,2}$ should be defined. In
MRK of hard processes~\cite{QMRK}, when initial-state radiation  is
not ordered in transverse momenta of emitted partons, transverse
momentum of initial-state parton in hard collision can be arbitrary
large up to some maximum value following from general kinematical
conditions, $t_{\infty}$.

When we study production of jets or associated production of photon
and jets, as in the case of present task, the upper limit for the
squares of the Reggeized parton's transverse momenta $t_1$ and $t_2$
should be truncated by the condition $t_1,t_2 < p_{2T}^2$, where
$p_{2T}^2$ is the smaller transverse momentum of a jet from the pair
of two leading jets. The above-mentioned condition arises from the
constraints of the jet-production experiment: one can measure an
azimuthal angle difference $\Delta\phi$ between the two most
energetic jets  but it is impossible to separate final-state partons
produced in the hard parton scattering phase from the ones generated
during the QCD evolution of unPDFs~\cite{NKS2017}. The MRK evolution
suggests a strong ordering in rapidity but the transverse momenta of
partons in the QCD ladder keep similar values. This means that the
transverse momenta of partons generated in the initial-state
evolution described via the unPDF must be smaller than the
transverse momenta of both measured leading jets,
$\sqrt{t_{1,2}}<p_{2T}<p_{1T}$. However, as in the relevant
experiment the leading and subleading jets are produced in some
central region of rapidity $|y^{1,2}|<Y$, there is a probability to
find partons with larger transverse momenta ($p_T>p_{2T}$),
originated from evolution of unPDFs outside this region of rapidity
and they can not be considered as leading jets, which should be in
central region of rapidity. Formally, we can rewrite integrals over
$t_{1,2}$  as sum of "DGLAP" (first term) and additional "MRK"
(second term) contributions:
\begin{equation}
\int\limits_0^{t_{\infty}}dt \Phi(x,t,\mu^2) \Rightarrow
\int\limits_{0}^{p_{2T}^2}dt \Phi(x,t,\mu^2) +
\int\limits_{p_{2T}^2}^{t_\infty}dt
w(x,t,\mu^2)\Phi(x,t,\mu^2),\label{eq:mrk}
\end{equation}
where $w(x,t,\mu^2)$ can be considered as damping function which is
independent on details of hard process but it can depend on
QCD-evolution of unPDF. In the first approximation, we can interpret
the average value $\overline{w}=\overline{w}(Y,p_{2T})$ as a
probability to find more energetic partons with rapidities $|y|>Y$
than ones defined as leading jets in the rapidity region
$|y^{1,2}|<Y$. After this we rewrite formulae (\ref{eq:mrk}) as
follows
\begin{equation}
\int\limits_0^{t_{\infty}}dt \Phi(x,t,\mu^2) \Rightarrow
\int\limits_{0}^{p_{2T}^2}dt \Phi(x,t,\mu^2) +
\overline{w}(Y,p_{2T})\times \int\limits_{p_{2T}^2}^{t_\infty}dt
\Phi(x,t,\mu^2).
\end{equation}
Further, we will consider $\overline{w}(Y,p_{2T})$ as a free
parameter of our model with two boundary conditions, which
correspond either DGLAP regime $(\overline{w}=0)$ or asymptotic MRK
regime $(\overline{w}=1)$. Such a way, the factorization formula
(\ref{eqI:kT_fact}) reads as
\begin{equation}
d\sigma=d\sigma^{DLAP}+\overline{w} d\sigma^{MRK1}+\overline{w}
d\sigma^{MRK2}+\overline{w}^2 d\sigma^{MRK12},
\end{equation}
where
\begin{eqnarray}
  d\sigma^{DGLAP} &=& \int\limits_0^1 \frac{dx_1}{x_1} \int \frac{d\phi_1}{2\pi}\int\limits_0^{p_{2T}^2} dt_1
  {\Phi}_1(x_1,t_1,\mu^2) \times \nonumber \\
& \times &\int\limits_0^1 \frac{dx_2}{x_2} \int
\frac{d\phi}{2\pi}\int\limits_0^{p_{2T}^2} dt_2
{\Phi}_2(x_2,t_2,\mu^2)\cdot d\hat{\sigma}^{\rm PRA},
\label{eq:DGLAP}
  \end{eqnarray}

\begin{eqnarray}
  d\sigma^{MRK1} &=& \int\limits_0^1 \frac{dx_1}{x_1} \int \frac{d\phi_1}{2\pi}\int\limits_{p_{2T}^2}^{t_\infty} dt_1
  {\Phi}_1(x_1,t_1,\mu^2) \times \nonumber \\
& \times &\int\limits_0^1 \frac{dx_2}{x_2} \int
\frac{d\phi}{2\pi}\int\limits_0^{p_{2T}^2} dt_2
{\Phi}_2(x_2,t_2,\mu^2)\cdot d\hat{\sigma}^{\rm PRA},
\label{eq:MRK1}
  \end{eqnarray}

\begin{eqnarray}
  d\sigma^{MRK2} &=& \int\limits_0^1 \frac{dx_1}{x_1} \int \frac{d\phi_1}{2\pi}\int\limits_0^{p_{2T}^2} dt_1
  {\Phi}_1(x_1,t_1,\mu^2) \times \nonumber \\
& \times &\int\limits_0^1 \frac{dx_2}{x_2} \int
\frac{d\phi}{2\pi}\int\limits_{p_{2T}^2}^{t_\infty} dt_2
{\Phi}_2(x_2,t_2,\mu^2)\cdot d\hat{\sigma}^{\rm PRA},
\label{eq:MRK2}
  \end{eqnarray}

\begin{eqnarray}
  d\sigma^{MRK12} &=& \int\limits_0^1 \frac{dx_1}{x_1} \int \frac{d\phi_1}{2\pi}\int\limits_{p_{2T}^2}^{t_\infty} dt_1
  {\Phi}_1(x_1,t_1,\mu^2) \times \nonumber \\
& \times &\int\limits_0^1 \frac{dx_2}{x_2} \int
\frac{d\phi}{2\pi}\int\limits_{p_{2T}^2}^{t_\infty} dt_2
{\Phi}_2(x_2,t_2,\mu^2)\cdot d\hat{\sigma}^{\rm PRA}.
\label{eq:MRK12}
  \end{eqnarray}

\section{Numerical results and discussion}

To study a relative role of MRK effects in azimuthal angle
decorrelation effects, we compute normalized $\Delta\phi$ spectra
for $\gamma+2jets$ events, which were measured by D0
Collaboration~\cite{D0gamma2} at the $\sqrt{S}=1.96$ TeV. Leading
and subleading jets are measured in the rapidity region
$|y^{1,2}|<Y=3.5$ while the photon rapidity belongs to the interval
$|y^\gamma|<2.5$, excluding subinterval $1.0<|y^\gamma|<1.5$.
Transverse momentum of photon is restricted by the condition
$50<p_T^{\gamma}<90$ GeV, and the leading jet has a transverse
momentum $p_{1T}>30$ GeV. The data were collected in three sets with
different conditions for subleading jet: set 1 -- $15<p_{2T}<20$
GeV, set 2 -- $20<p_{2T}<25$ GeV, set 3 -- $25<p_{2T}<30$ GeV.

\begin{figure}[h]
\centering
\includegraphics[width=0.8\textwidth, angle=0]{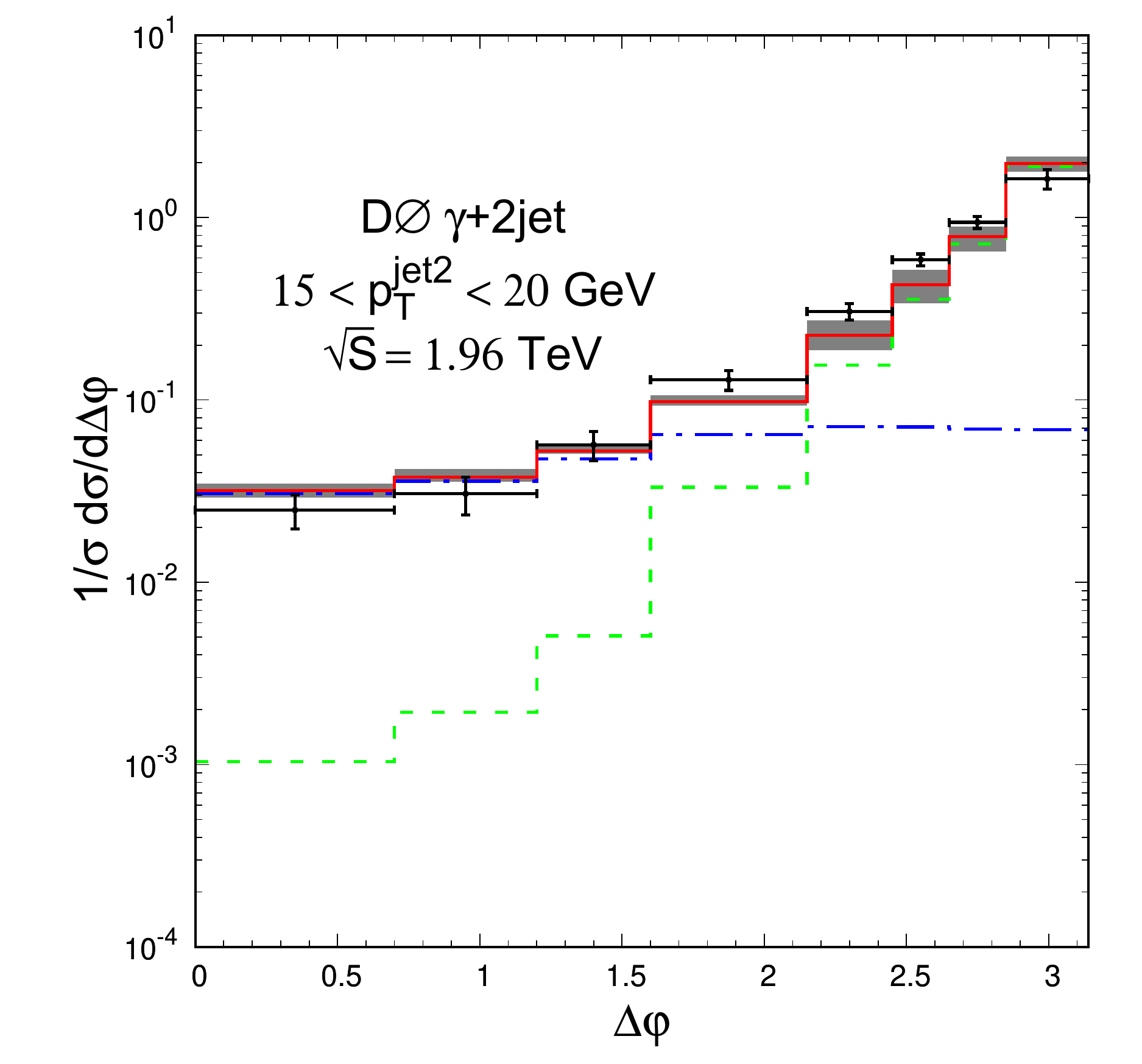}
\caption{Azimuthal angle difference $(\Delta\phi)$ spectrum  at the
$15<p_{T}^{jet2}<20$ GeV. The dashed green line is DGLAP
contribution, the dash-dotted blue line is MRK contribution with
optimized $\overline{w}$ from the Table~\ref{TableII} and the solid
red line is their sum. The data are from D0
Collaboration~\cite{D0gamma2}.}
\label{fig-1}       
\end{figure}

\begin{figure}[h]
\centering
\includegraphics[width=0.8\textwidth, angle=0]{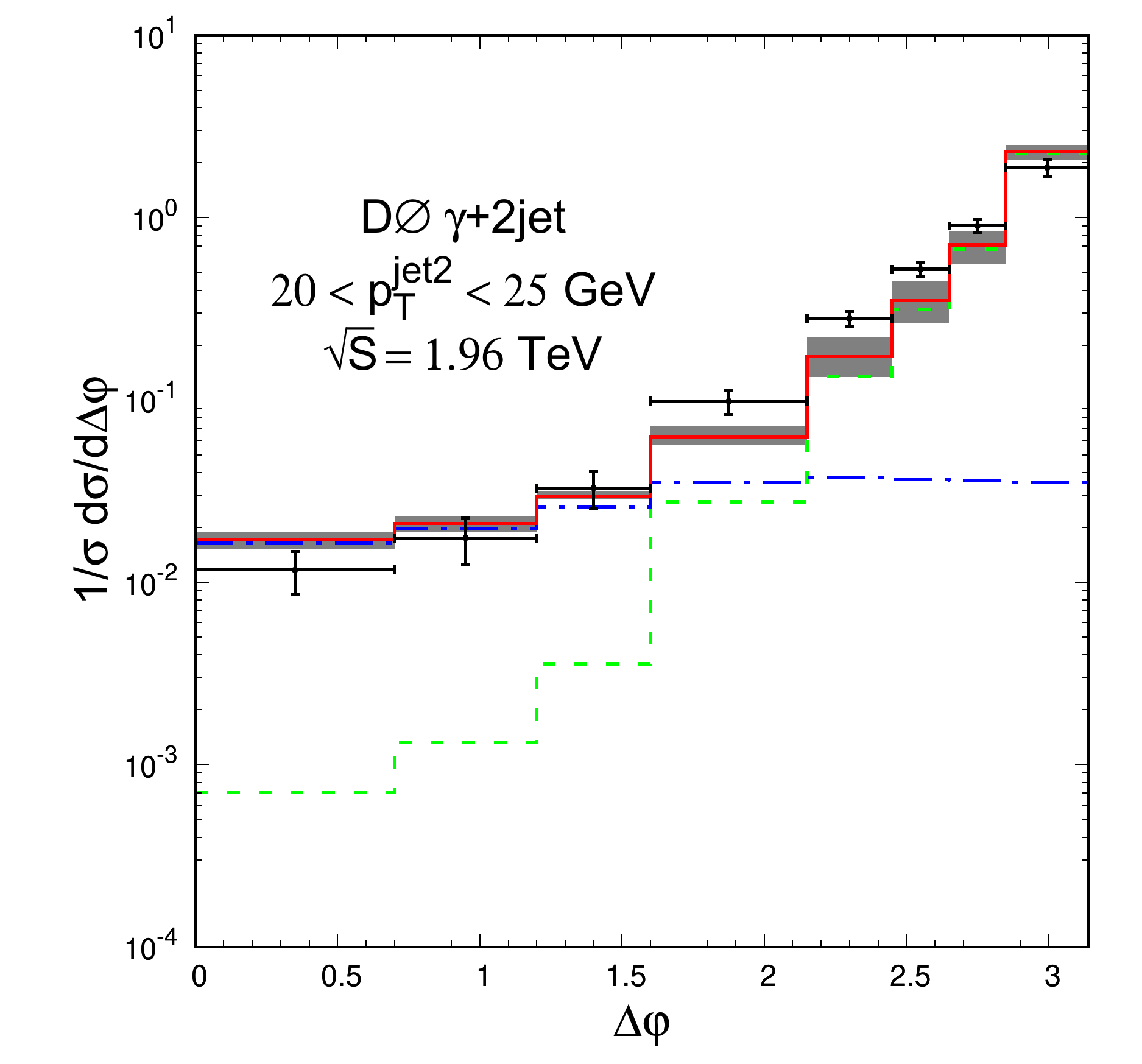}
\caption{Azimuthal angle difference $(\Delta\phi)$ spectrum  at the
$20<p_{T}^{jet2}<25$ GeV. The curves are defined as in the Fig.~\ref{fig-1}. The
data are from D0 Collaboration~\cite{D0gamma2}.}
\label{fig-2}       
\end{figure}

\begin{figure}[h]
\centering
\includegraphics[width=0.8\textwidth, angle=0]{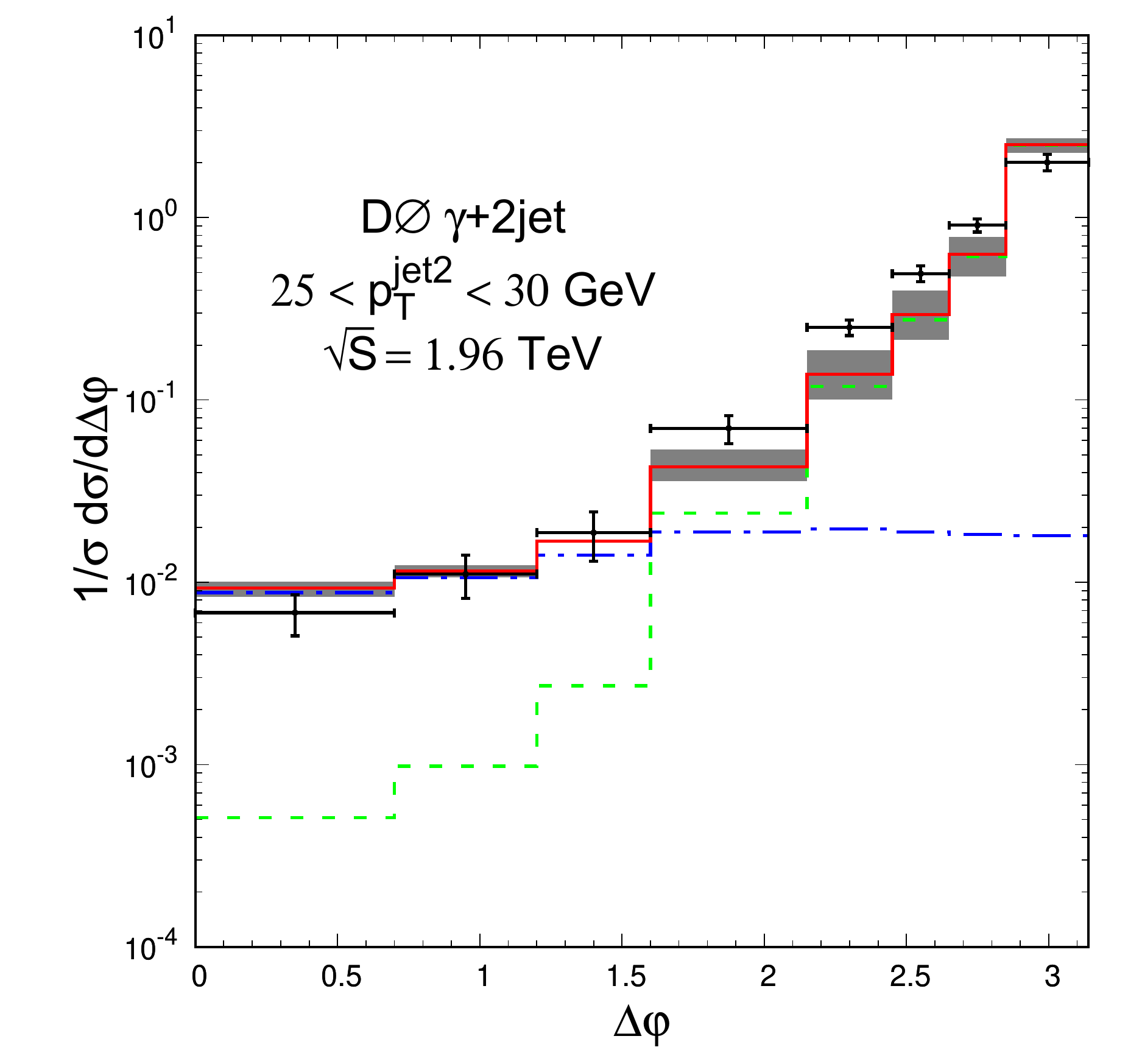}
\caption{Azimuthal angle difference $(\Delta\phi)$ spectrum  at the
$25<p_{T}^{jet2}<30$ GeV. The curves are defined as in the Fig.~\ref{fig-1}. The
data are from D0 Collaboration~\cite{D0gamma2}.}
\label{fig-3}       
\end{figure}

The results of calculation in DGLAP approximation (\ref{eq:DGLAP})
for azimuthal angle difference spectra are shown in Fig.~\ref{fig-1}
- \ref{fig-3} as dashed green lines. The same as NLO calculations in
CPM~\cite{SHERPA}, they strongly underestimate data at the small
$\Delta\phi$ and the inclusion of DPS contributions is needed.
Following Ref.\cite{D0gamma2}, we define $\beta$ as a fraction of DPS
events:
\begin{equation}
\frac{1}{\sigma}\frac{d\sigma}{d\Delta
\phi}=(1-\beta)\frac{1}{\sigma^{SPS}}\frac{d\sigma^{SPS}}{d\Delta
\phi} +\beta \frac{1}{\sigma^{DPS}}\frac{d\sigma^{DPS}}{d\Delta
\phi}
\end{equation}
Our estimations for parameters $\beta^{PRA}$ in calculations using
DGLAP approximation at the different values of $p_{2T}$ are
presented in the Table ~\ref{TableI} as well as the values of
$\beta^{SHERPA}$ obtained in NLO CPM calculations. Both
parameters are extracted by a fit of experimental data in assumption
that DPS contribution is constant throughout the $\Delta\phi$-spectra. We find
a good agreement between NLO collinear and LO $k_T$-factorized
schemes of calculations based on DGLAP picture of initial-state QCD
evolution, proving the earlier obtained results for many other
processes.

At very large energy in the MRK limit, when QCD evolution of
$t-$channel partons should be described by BFKL equations
\cite{BFKL} instead of DGLAP equations~\cite{DGLAP}, the transverse
    momenta of Reggeized partons connecting gauge-invariant
    clusters can be arbitrary large and independent on transverse
momenta of final particles in clusters.

At the present energy such MRK picture can be realized in some part
and we suggest to control a signal of MRK regime adding to LO PRA
cross section the $d\sigma^{MRK1,2}$ and $d\sigma^{MRK12}$ terms.
The phenomenological parameter $\overline{w}$ can be extracted from
the data, see Table~\ref{TableII}. In the
Figs.~\ref{fig-1}-\ref{fig-3}, the normalized azimuthal angle
difference spectra at the different cuts on transverse momentum of
subleading jet $p_{2T}$ are shown. We found the MRK contribution to
the total cross section to be small,
\begin{equation}
R(p_{2T})=\frac{\sigma^{MRK}}{\sigma^{DGLAP}+\sigma^{MRK}}=0.046
\div 0.163.\label{Rpt2}
\end{equation}
We see that probability $\overline{w}$ degenerates with growth of
$p_{2T}$ following our guess, as well as MRK contribution to the
total cross section, see Table~\ref{TableII}. As it can be expected,
initial-state partons with large transverse momenta contribute to
the region of small $\Delta\phi$ enhancing decorrelation effect.

Finally, we would like to note that a theoretical calculation of
modified unPDFs, which should be dependent on transverse momenta of
leading hard jets and the rapidity region $Y$ where they are
measured, can be implemented principally, providing an additional
test for the already known models of unPDFs, such as used here
KMR~\cite{KMR} as well as the ones collected in TMD-library
\cite{TMDlib}, and also recently suggested unPDFs based on Parton
Branching method~\cite{PBM}.

\section*{Acknowledgements}
We are grateful to Maxim Nefedov for help in the stage of analytical
calculations with model-file ReggeQCD. Authors thank the Ministry of
Education and Science of the Russian Federation for financial
support in the framework of the Samara University Competitiveness
Improvement Program among the world's leading research and
educational centers for 2013-2020, the task number 3.5093.2017/8.9.
Authors (A.K. and V.S.) thank the Foundation for the Advancement of
Theoretical Physics and Mathematics BASIS, grant No. 18-1-1-30-1.

\newpage

\begin{table}[h]
 \tbl{The fraction of DPS events $\beta$. PRA results include only DGLAP contribution (\ref{eq:DGLAP}).}
 {\begin{tabular}{@{}ccc@{}}
 \Hline
 \\[-1.8ex]
 $p_{2T}$, GeV & $\beta^{PRA}$, \% & $\beta^{SHERPA}$, \% \\
 \hline
 $15-20$ & $10.4^{+1.1}_{-1.0}$ & $11.6\pm 1.4$ \\
 \hline
 $20-25$ & $5.1^{+0.6}_{-0.7}$  & $5.0\pm 1.2$ \\
 \hline
 $25-30$ & $2.7^{+0.3}_{-0.4}$ & $2.2 \pm 0.8$ \\
\Hline \\[-1.8ex]
 \end{tabular}}
\label{TableI}
 \end{table}

\begin{table}[h]
 \tbl{The MRK parameter $\overline{w}$ and cross-section ratio $R$ (\ref{Rpt2}) calculated in the LO PRA.}
{ \begin{tabular}{@{}ccc@{}}
 \Hline
 \\[-1.8ex]
 $p_{2T}$, GeV & $\overline{w}$, \% & $R$, \% \\
 \hline
 $15-20$ & $24^{+17}_{-11}$ & $16^{+8}_{-4}$  \\
\hline
 $20-25$ & $17^{+11}_{-8}$ & $9^{+9}_{-6}$\\
\hline
 $25-30$ & $11^{+7}_{-5}$ & $5^{+4}_{-3}$ \\
 \Hline \\[-1.8ex]
 \end{tabular}}
\label{TableII}
 \end{table}

 \newpage

\begin{thebibliography}{}
\bibitem{DPS} M.~Diehl and J.~R.~Gaunt, arXiv:1710.04408 [hep-ph].

\bibitem{D0gamma2} V. M. Abazov et al. [D0 Collaboration], {\it Phys. Rev.} \textbf{D93}, 052008 (2016) .
\bibitem{NSS2013} M.A. Nefedov, V.A. Saleev, A.V. Shipilova, {\it  Phys. Rev.}  \textbf{D87}, 094030 (2013).

\bibitem{gammaHERA}   B. A. Kniehl, M. A. Nefedov, and V. A. Saleev, {\it Phys. Rev.} \textbf{D89}, 114016 (2014).

\bibitem{SHERPA} T. Gleisberg et al., {\it JHEP} {\bf 0902}, 007
(2009).

\bibitem{NKS2017} A.V. Karpishkov, M.A. Nefedov,  V.A. Saleev,
{\it Phys. Rev.} \textbf{D96}, 096019 (2017).


\bibitem{PRANLO1} M.A. Nefedov, V.A. Saleev, {\it Mod.Phys.Lett.} \textbf{A32}, 1750207 (2017).

\bibitem{KMR} M.A. Kimber, A.D. Martin, M.G. Ryskin, {\it Phys. Rev.} \textbf{D63}, 114027 (2001).

\bibitem{Lipatov95} L.N. Lipatov, {\it Nucl. Phys.} \textbf{B452}, 369 (1995).

\bibitem{AntonovLipatov} E.N. Antonov, L.N. Lipatov, E.A. Kuraev, I.O.
Cherednikov, {\it  Nucl. Phys.} \textbf{B721}, 111 (2005).

\bibitem{LipatovVyazovsky} L.N. Lipatov, M.I. Vyazovsky, {\it
 Nucl. Phys.} \textbf{B597}, 399 (2001).


\bibitem{gamma} V.A.  Saleev, {\it Phys.Rev.} \textbf{D78}, 034033 (2008).

\bibitem{gamma-jet} B.A. Kniehl, V.A. Saleev, A.V. Shipilova, E.V. Yatsenko, {\it Phys. Rev.} \textbf{D84}, 074017
(2011).
\bibitem{Diphotons} M.A. Nefedov, V.A.  Saleev, {\it Phys. Rev.} \textbf{D92}, 094033 (2015).

\bibitem{ReggeQCD} M.A. Nefedov and V.A. Saleev, {\it Phys. Rev.} \textbf{D92}, 094033 (2015).

\bibitem{FeynArts} T. Hahn, {\it Comput. Phys. Commun.} \textbf{140}, 418 (2001).

\bibitem{MRST}  A. D. Martin, W. J. Stirling, and R. S. Thorne, {\it Phys. Lett. B} \textbf{636},
259 (2006).

 \bibitem{Katie}   A.~van Hameren,
  {\it  Comput.Phys.Commun.} {\bf 224}, 371 (2018).

\bibitem{Hameren1} A. van Hameren, P. Kotko, K. Kutak. Helicity amplitudes for high-energy scattering, {\it
JHEP}
{\bf 01}, 078 (2013).
\bibitem{Hameren2} A. van Hameren, K. Kutak, T.
Salwa, {\it Phys. Lett.} {\bf B727}, 226 (2013).




\bibitem{QMRK}  
  V.~S.~Fadin and L.~N.~Lipatov,
  {\it Nucl.\ Phys.}  {\bf B406}, 259 (1993).


\bibitem{DGLAP} V.~N.~Gribov and L.~N.~Lipatov,
{\it Sov.\ J.\ Nucl.\ Phys. } \textbf{15}, 438 (1972);
Yu.~L.~Dokshitzer, Sov.\ Phys.\ {\it JETP} \textbf{46}, 641 (1977);
G.~Altarelli and G.~Parisi, {\it Nucl.\ Phys.} \textbf{B126}, 298
(1977).



%
\bibitem{BFKL} 
  L.~N.~Lipatov,
 {\it  Sov.\ J.\ Nucl.\ Phys.}  {\bf 23}, 338 (1976);
  E.~A.~Kuraev, L.~N.~Lipatov, and V.~S.~Fadin,
  {\it Sov.\ Phys. JETP} {\bf 44}, 443 (1976);
  {\it Sov.\ Phys.\ JETP} {\bf 45}, 199 (1977);
  I.~I.~Balitsky and L.~N.~Lipatov,
 {\it  Sov.\ J.\ Nucl.\ Phys. } \textbf{ 28}, 822 (1978);
 {\it Sov.\ Phys.\ JETP} \textbf{ 63}, 904 (1986).

\bibitem{TMDlib}  F. Hautmann et al., {\it Eur. Phys. J.} \textbf{C74}, 3220 (2014).

\bibitem{PBM} A.~Bermudez Martinez, P.~Connor, F.~Hautmann, H.~Jung, A.~Lelek, V.~Radescu and R.~Zlebcik,
  arXiv:1804.11152 [hep-ph].
\end{thebibliography}
\end{document}